\documentclass[twocolumn,twoside,slac_two]{revtex4}
\usepackage{graphicx}
\usepackage{fancyhdr}
\usepackage{graphics}
\usepackage{epstopdf}
\usepackage{textpos}
  \newcommand{\GeV}{\ensuremath{\,\text{GeV}}}
  \newcommand{\MET}{\ensuremath{E_T^{\text{miss}}}}
  \newcommand{\ttbar}{\ensuremath{t\bar{t}}}
  \newcommand{\Z}{\ensuremath{Z^0}}
  \newcommand{\W}{\ensuremath{W^\pm}}
  \newcommand{\HT}{\ensuremath{H_T}}
  \newcommand{\pT}{\ensuremath{p_T}}

  \newcommand{\mZ}{\ensuremath{m_0}}
  \newcommand{\mH}{\ensuremath{m_{1/2}}}

  \newcommand{\chiZ}{\ensuremath{\tilde{\chi}^0_2}}
  \newcommand{\LSP}{\ensuremath{\tilde{\chi}^0_1}}
\begin{document}

\title{\centering SUSY Searches with Leptons in the Final State at CMS}
\author{
\centering
\begin{center}
Hannes Helmut Schettler for the CMS Collaboration
\end{center}}
\affiliation{\centering DESY / University of Hamburg, Germany}
\begin{abstract}
Recent results of the searches for Supersymmetry in final states with one or two leptons at CMS are presented. 
Many Supersymmetry scenarios, including the Constrained Minimal Supersymmetric extension of the Standard Model (CMSSM), predict a substantial amount of events containing leptons, while the largest fraction of Standard Model background events -- which are QCD interactions -- gets strongly reduced by requiring isolated leptons. 
The analyzed data was taken in 2011 and corresponds to an integrated luminosity of approximately $\mathcal{L} = 1\,\text{fb}^{-1}$. The center-of-mass energy of the pp collisions was $\sqrt{s}=7\,\text{TeV}$.
\end{abstract}

\maketitle
\thispagestyle{fancy}


\section{Introduction}
Supersymmetry (SUSY) is a well motivated extension of the Standard Model (SM) (see e.\,g.~\cite{martin}). 
Several problems of the SM could be solved by SUSY, such as the hierarchy problem in the Higgs sector. 
Furthermore, many SUSY models naturally provide a Dark Matter candidate.

The presented analyses are based on $1\,\text{fb}^{-1}$ of data taken at the Compact Muon Solenoid (CMS) at the Large Hadron Collider (LHC). A description of the detector can be found elsewhere~\cite{cms}.

Signatures with leptons, missing transverse energy (\MET), and hadronic activity (i.\,e.\ jets) are promising, as outlined below.

In pp collisions colored particles are predominantly produced.
Colored SUSY particles can decay into jets and other SUSY particles, which may be color-neutral.
In a second decay step the latter ones may decay into leptons and SUSY particles.
Even if the decay into leptons is not dominant, selecting events with leptons can still be suitable since the SM background very effectively gets reduced.
Assuming $R$ parity conservation, a SUSY particle never solely decays to SM particles, and the decay chain ends with the lightest SUSY particle (LSP), which can be assumed to be undetectable, causing an \MET\ signature in the event.

In the following sections an overview of three analyses with differing leptonic content of the events is presented: 
exactly one lepton, at least two leptons with opposite charge, and at least two leptons with the same charge. The complete CMS publications can be found in \cite{SL_PAS}, \cite{OS_PAS}, and \cite{SS_PAS}.

\section{Single Lepton Channel}\label{SL}
In this channel events with exactly one isolated electron or muon are considered.
In background events single leptons typically originate from a leptonic \W\ decay, either directly produced, or in \ttbar\ events. The other decay product of the \W\ is a neutrino causing \MET\ in the events.
Therefore, in SM events a relation is expected between \MET\ and the charged lepton.

\begin{figure}
  \centering
  \includegraphics[width=3.9cm]{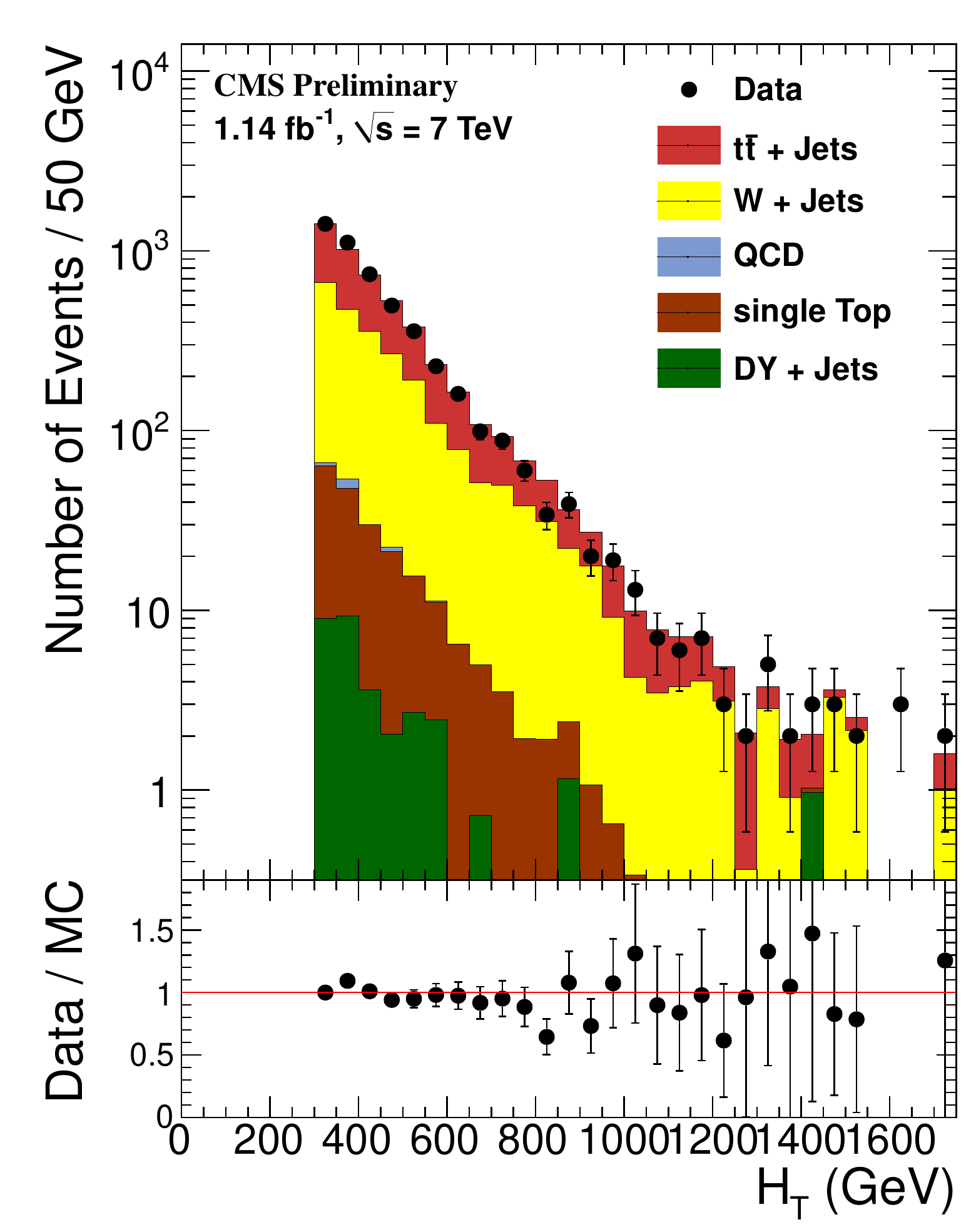}
  \includegraphics[width=3.9cm]{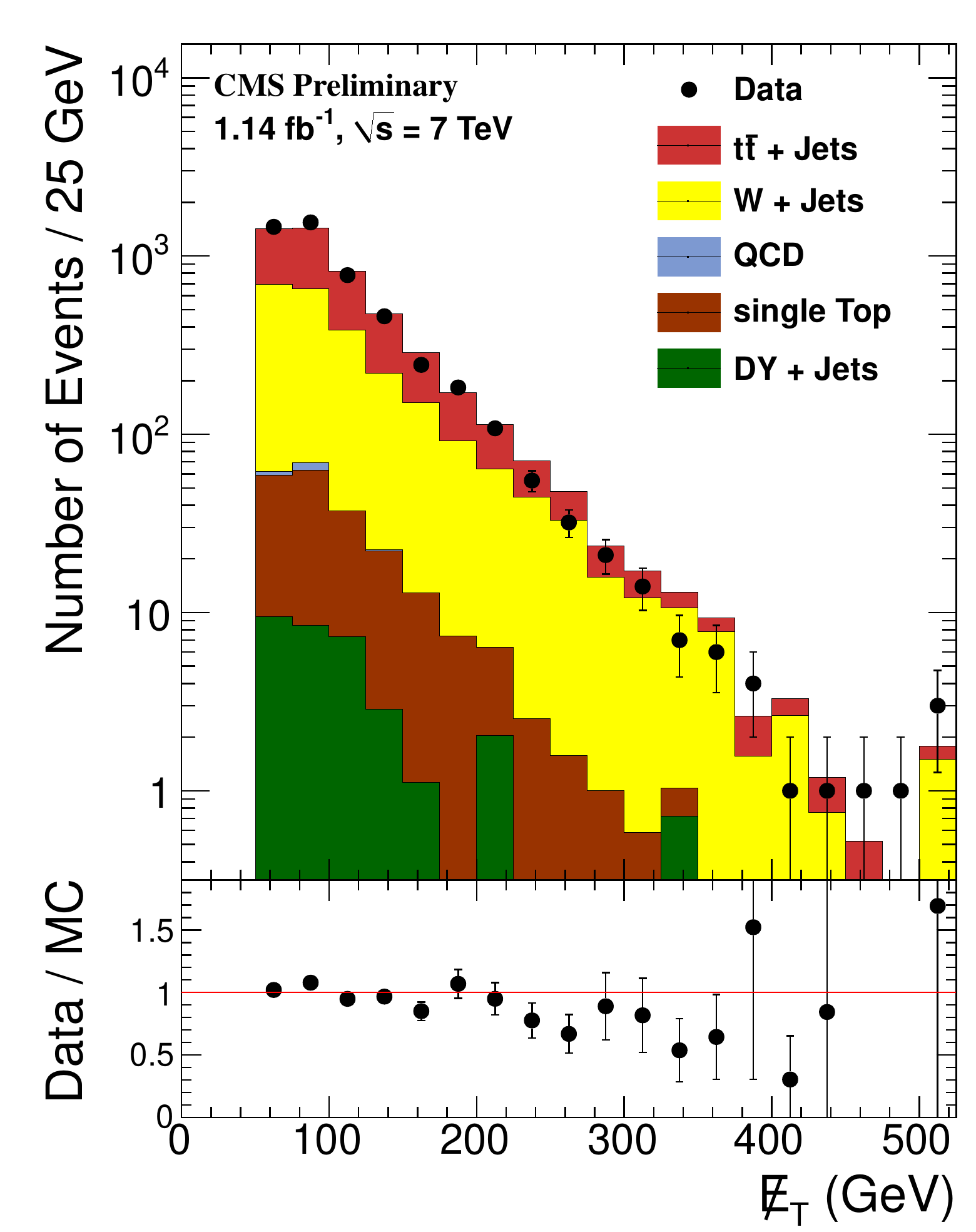}
\caption{Distributions of $H_T$ and \MET\ in the single $\mu$ channel comparing data and simulation.\label{SL_control}}
\end{figure}
Events are selected based on a trigger requiring one lepton and hadronic activity which is measured with $\HT=\sum p_T^{\text{jet}}$.
In the offline selection one electron or muon with $\pT>20\GeV$ and an \HT\ larger than $300\GeV$ is required.
Furthermore, only events with at least three jets and $\MET>60\GeV$ are selected. 
Figure~\ref{SL_control} shows the distributions of \MET\ and \HT\ after this selection.
As expected, the background consists mostly of \ttbar\ and \W\ events.

Two methods were developed using the fact that \MET\ and the momentum of the charged lepton is related in background but not in the expected signal.

\subsection{Lepton Spectrum Analysis}
The basic idea of the lepton spectrum method is to use the spectrum of the charged lepton to predict the \MET\ spectrum.
The \MET\ in most SUSY models is dominated by two LSPs in the event. Hence, it is uncorrelated to the lepton \pT\ and its spectrum extends to higher values.

Three issues were addressed to make this prediction reliable:
the effect of \W\ polarization, the bias of the applied lepton \pT\ cut, and the
difference in resolution on lepton \pT\ and \MET\ measurements.

The background from dilepton processes (including $\tau$) is estimated by simulating the loss of one lepton in data events with two reconstructed leptons.

\subsection{Lepton Projection Method}
Here, the correlation of the charged lepton and the neutrino caused by the $V-A$ nature of the \W\ decay is used to predict the main background.

An angular analysis of the decay products in the \W\ rest frame would exhibit a significant asymmetry between the charged lepton and the neutrino.
Since the momentum of the neutrino along the beam axis cannot be measured, an observable is used, which depends only on transverse components:
\begin{equation}
L_P = \frac{\vec{p}_T(\ell)\cdot\vec{p}_T(W)}{|\vec{p}_T(W)|^2}
\end{equation}
The SM background occupies a broad range of $L_P$ values, while SUSY signatures are expected to peak at zero.
In a SM dominated control region ($L_P>0.3$) MC templates are fitted to data.
For the prediction an extrapolation is performed in the signal region ($L_P<0.15$).

\begin{figure}
\centering
  \includegraphics[width=7cm]{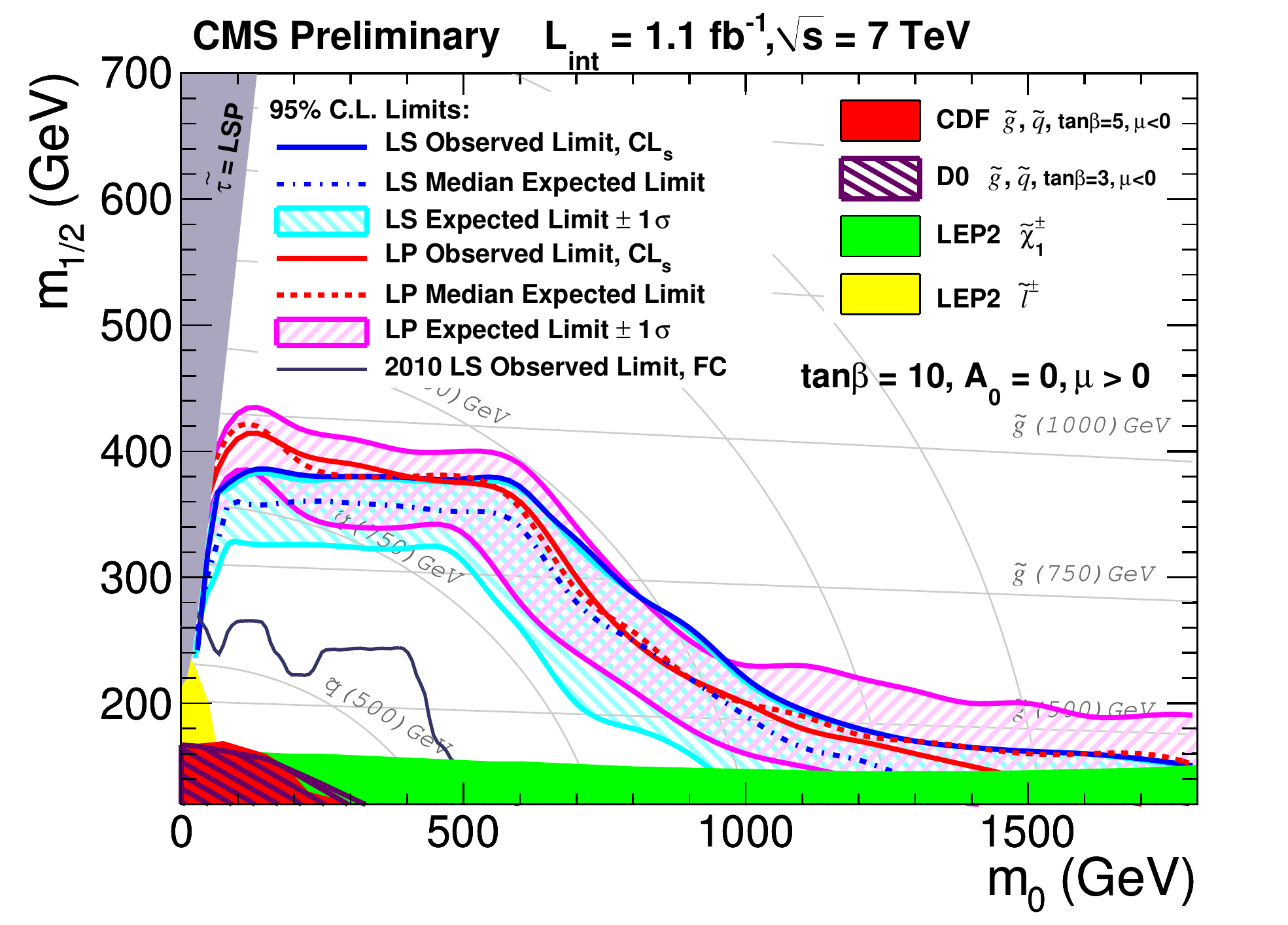}
\caption{Exclusion region in the CMSSM \mZ-\mH\ plane, shown for both the lepton spectrum method (LS) and the lepton projection variable method (LP).\label{SL_res}}
\end{figure}

\subsection{Result}
The results of both methods are compatible with the expectations from SM backgrounds.
Figure~\ref{SL_res} shows the parameter regions in the CMSSM that are excluded by these analyses.

\section{Dilepton Opposite-Sign Channel}\label{OS}
The dilepton opposite-sign analysis consists of two approaches.
The first one makes use of a typical decay in SUSY: $\chiZ\rightarrow\ell\tilde{\ell}\rightarrow\LSP\ell^+\ell^-$.
In the dilepton mass spectrum this decay leads to a mass edge. Therefore the shape of the distribution is analyzed.
The second part of the analysis is a counting experiment.

After triggering on two leptons, two isolated opposite-sign leptons are selected with $p_T>20\GeV$ for the harder one and $p_T>10\GeV$ for the softer. 
Events with a same-flavor pair in the mass regions between $76\GeV$ and $106\GeV$ or below $12\GeV$ are rejected. 
Furthermore, at least two jets are required, as well as $\HT>100\GeV$, and $\MET>50\GeV$.

\begin{figure}[b]
\centering
  \includegraphics[width=6cm]{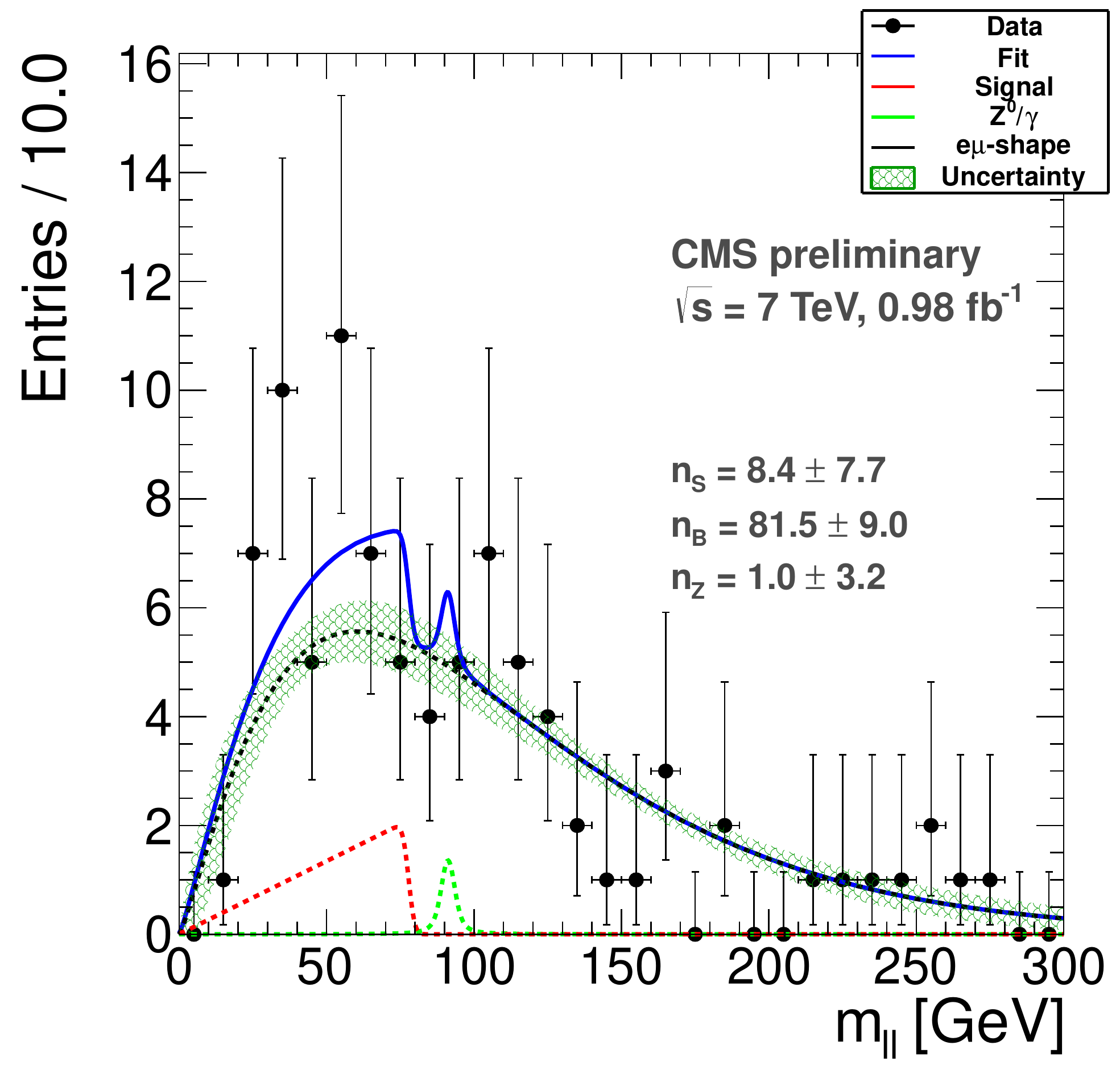}
\caption{Opposite-sign same-flavor dilepton mass with a combined fit of background (modeled by opposite-flavor events), the \Z peak, and a possible signal.\label{OS_edge}}
\end{figure}

\subsection{Search for Kinematic Edge}
Many SUSY models lead to characteristic edges in the dilepton mass spectrum.
Kinematically there is an upper bound on the energy that is carried away from the leptons, which is mainly driven by the mass difference of the initial SUSY particle and the LSP.
Assuming the lepton flavor to be conserved, the edge is only expected in the same-flavor spectrum.
In the dominant background process \ttbar\ the two lepton flavors are uncorrelated and the spectrum in the opposite-flavor channel can be used to predict the SM yield in the same-flavor channel.

Figure~\ref{OS_edge} shows the result. In addition to the SM background an insignificant signal contribution, modeled by a Gaussian smeared triangle, can be fitted.
\subsection{Counting Experiment}
For the counting experiment 
two signal regions are defined:
a high \MET\ region ($\MET>275\GeV$, $\HT>300\GeV$) and a high \HT\ region ($\MET>200\GeV$, $\HT>600\GeV$).
The background is predicted in a data-driven way via two methods.

The first one is a factorization method which accounts for the weak correlation of the two used variables \HT\ and $y=\MET/\sqrt{\HT}$.
In the second method the \MET\ spectrum is modeled with the dilepton spectrum, corrected for the effects of the \W\ polarization.

\begin{figure}[!t]
\centering
  \includegraphics[width=7cm]{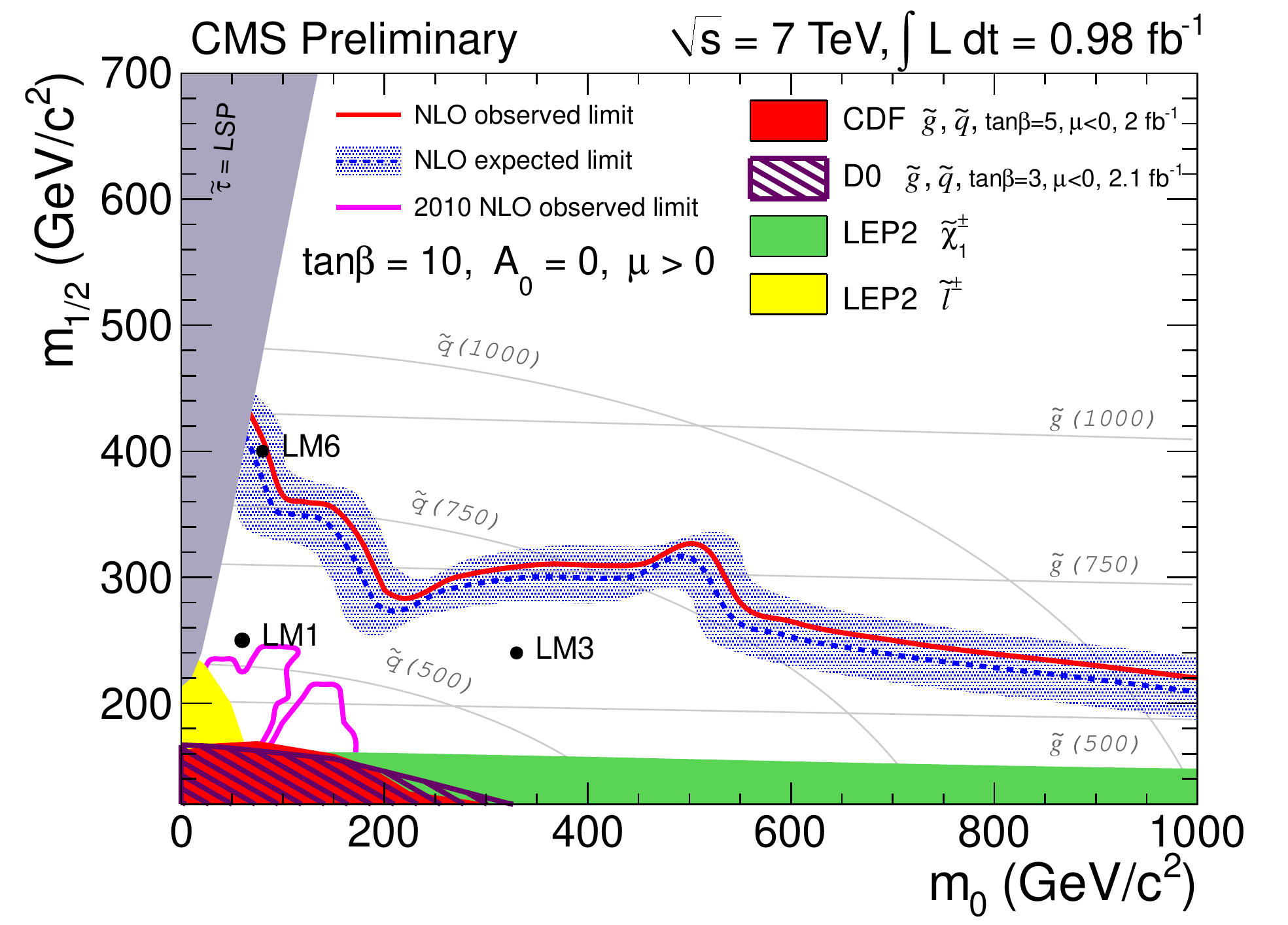}
\caption{Exclusion region in the CMSSM \mZ-\mH\ plane gained by the dilepton opposite-sign analysis.\label{OS_res}}
\end{figure}

\subsection{Result}
Both the results of the search for a kinematic edge and the counting experiment are compatible with the SM-only hypothesis. Within the CMSSM the parameter space shown in Fig.~\ref{OS_res} can be excluded.

\section{Dilepton Same-Sign Channel}\label{SS}
Two isolated leptons with the same charge are very rare in SM events, but could naturally appear in the decay chains of SUSY processes.

There are two selection strategies chosen for this analysis. One is based on dilepton triggers. One of the two same-sign leptons must exceed $20\GeV$ in \pT, the other $10\GeV$, \HT\ is required to be larger than $80\GeV$. 
The other selection includes very soft leptons ($\pT^\mu>5\GeV$, $\pT^e>10\GeV$) and requires $\HT>200\GeV$. 
The latter cut is necessary since a dilepton \HT\ cross-trigger is used.
In both cases only events with at least two jets and $\MET>30\GeV$ are considered.

\subsection{Backgrounds}
SM processes containing a pair of same-sign leptons, such as diboson or \ttbar\W\
production, have small cross sections and their contributions are predicted by simulation.
Most background is expected from events where one of the leptons is a so called {\it fake} lepton, especially semi-leptonic \ttbar\ with one lepton originating from a $b$ decay.

For the background estimation several methods are used: 
In a tight-to-loose method the ratio of leptons passing a tight selection and leptons passing a loose selection is measured in a multijet dominated control region. 
By applying this ratio on the loose leptons in the signal region, the number of fake leptons fulfilling the tight criteria can be estimated.

In a factorization method the isolation requirement for each lepton and the \MET\ requirement is assumed to be uncorrelated.
Then, the background from two fake leptons can be estimated by multiplying the probabilities for each lepton to pass the isolation cut and the event to pass the \MET\ cut.

In a third method the distribution of the isolation variable is measured in $b\bar{b}$ events and reweighted to the kinematics in the signal region.
From this the number of leptons originating from $b$ quarks and fulfilling the isolation requirement can be estimated.

\subsection{Result}
\begin{figure}[!h]
\centering
  \includegraphics[width=7cm]{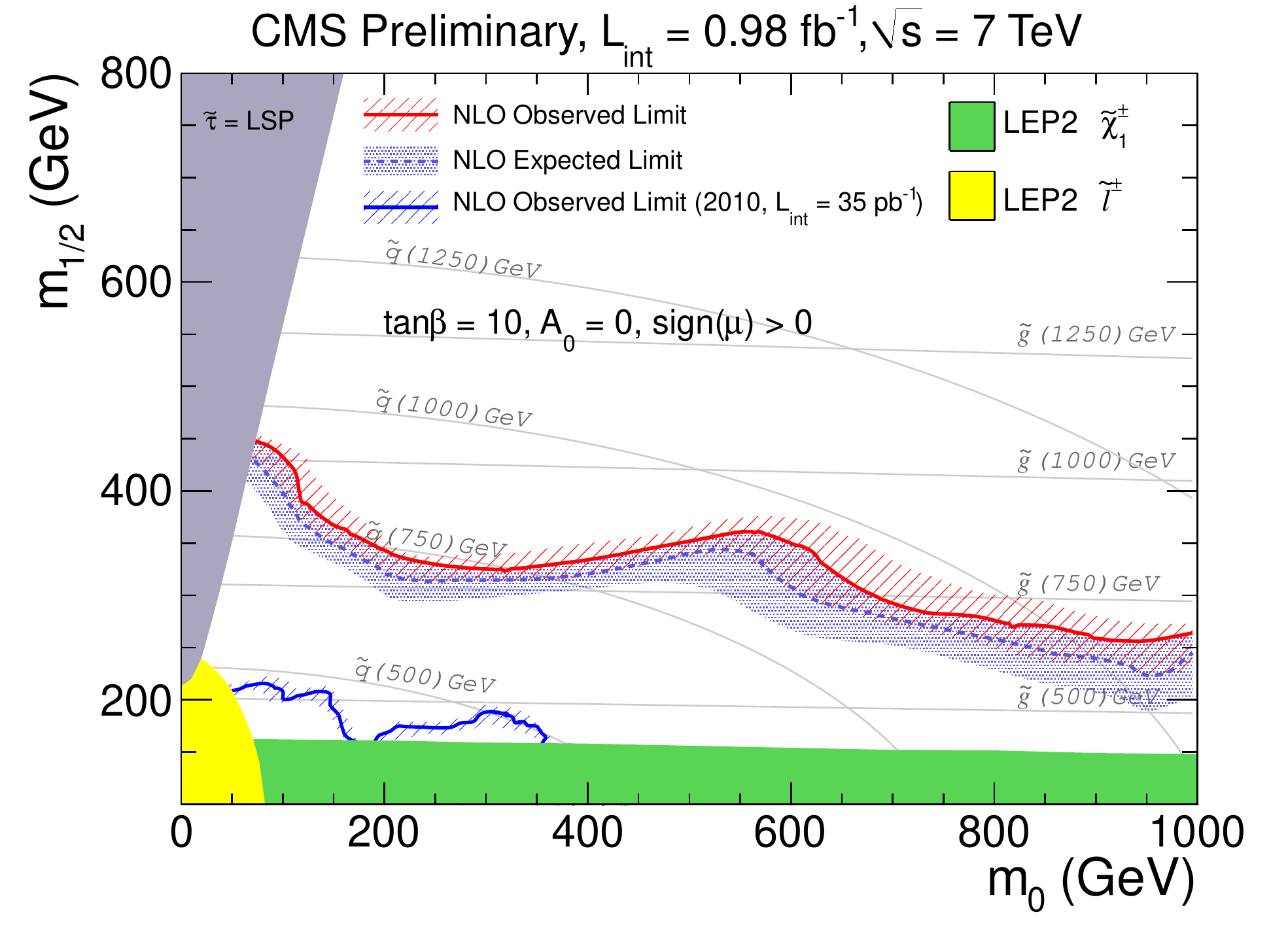}
\caption{Exclusion region in the CMSSM \mZ-\mH\ plane gained by the dilepton same-sign analysis.\label{SS_res}}
\end{figure}

All background estimation methods gain numbers which agree with SM. Figure~\ref{SS_res} shows the excluded area in CMSSM parameter space.

\section{Conclusion}
In none of the presented leptonic SUSY analyses a signal can be found in the first $1\,\text{fb}^{-1}$ of LHC data.
The exclusion limits in the parameter space of the CMSSM by far surpass the previous limits.



\bigskip 
\bibliography{basename of .bib file}

\end{document}